\documentclass[3p]{elsarticle}

\usepackage{graphicx}
\usepackage[colorlinks=true]{hyperref}
\usepackage{amssymb}
\usepackage{siunitx}
\usepackage{upgreek} 
\usepackage{color} 
\definecolor{blue}{rgb}{0,0,0.8}

\definecolor{red}{rgb}{0.8,0,0}

\definecolor{green}{rgb}{0,0.4,0}

\definecolor{violet}{rgb}{0.8,0.2,0.8}
\definecolor{darkgreen}{RGB}{9, 90, 6}

\usepackage[normalem]{ulem}
\usepackage{multirow}
\usepackage{multicol}
\usepackage{nonfloat}

\definecolor{darkPurple}{RGB}{150,0,150}
\definecolor{lightBrown}{rgb}{0.7,0.7,0.5}

\usepackage{ifthen}
\newboolean{clean}
\setboolean{clean}{true}  

\ifthenelse{\boolean{clean}}
{
    \newcommand\commentOld[1]{\bigskip}
    \newcommand\commentNew[1]{#1}
}
{
    \newcommand\commentOld[1]{{\color{lightBrown}#1}}
    \newcommand\commentNew[1]{{\color{darkPurple}#1}}
}

\usepackage{amsmath}
\usepackage{lineno}
\usepackage[symbol]{footmisc}

\biboptions{authoryear}
\journal{NeuroImage}
\usepackage{geometry}
\usepackage{lscape}
\usepackage{array}
\usepackage{booktabs}
\usepackage{hyperref}
\usepackage{tikz}
\usepackage{marginnote}
\newcommand{\leftmarginnote}[1]{
  \ifthenelse{\boolean{clean}}%
    {} 
    {
      \reversemarginpar 
      \marginnote{\tikz[baseline=(X.base)]\node [draw, fill=yellow, text width=1.5cm, align=center, rounded corners] (X) {#1};}
      \normalmarginpar
    }
}
\newcommand{\rightmarginnote}[1]{
  \ifthenelse{\boolean{clean}}%
    {} 
    {
      \marginnote{\tikz[baseline=(X.base)]\node [draw, fill=yellow, text width=1.5cm, align=center, rounded corners] (X) {#1};}
    }
}

\journal{}
\makeatletter
\def\ps@pprintTitle{}
\makeatother

\begin{document}
\begin{frontmatter}

\title{Mapping \commentNew{tissue microstructure of brain white matter} in vivo in health and disease using diffusion MRI}
\leftmarginnote{Address Comment R3.2} 

\author[1,2]{Ying Liao\corref{cor1}}\ead{ying.liao@nyulangone.org}
\author[1,2]{Santiago Coelho}
\author[1,2]{Jenny Chen}
\author[1,2]{Benjamin Ades-Aron}
\author[3]{Michelle Pang}
\author[1,2]{Valentin Stepanov}
\author[4]{Ricardo Osorio}
\author[1,2]{Timothy Shepherd}
\author[1,2]{Yvonne W. Lui}
\author[1,2]{Dmitry S. Novikov}
\author[1,2]{Els Fieremans}
\cortext[cor1]{Corresponding author}

\address[1]{Bernard and Irene Schwartz Center for Biomedical Imaging, Department of Radiology, New York University Grossman School of Medicine, New York, 10016, NY, USA}
\address[2]{Center for Advanced Imaging Innovation and Research (CAI$^2$R), Department of Radiology, New York University Grossman School of Medicine, New York, 10016, NY, USA}
\address[3]{John A. Burns School of Medicine, University of Hawaii at Manoa, Honolulu, 96813, HI, USA}
\address[4]{Department of Psychiatry, New York University Grossman School of Medicine, New York, 10016, New York, USA}

\begin{abstract}
Diffusion magnetic resonance imaging offers unique \textit{in vivo} sensitivity to tissue microstructure in brain white matter, which undergoes significant changes during development and is compromised in virtually every neurological disorder. Yet, the challenge is to develop biomarkers that are specific to micrometer-scale cellular features in a human MRI scan of a few minutes. Here we quantify the sensitivity and specificity of a multicompartment diffusion modeling framework to the density, orientation and integrity of axons. We demonstrate that using a machine learning based estimator, our biophysical model captures the morphological changes of axons in early development, acute ischemia and multiple sclerosis (total N=821). The methodology of microstructure mapping is widely applicable in clinical settings and in large imaging consortium data to study development, aging and pathology.
\end{abstract}


\end{frontmatter}

\begin{multicols}{2}
\section{Introduction}
Diffusion magnetic resonance imaging (dMRI) is a non-invasive technique that maps the probability density $S(t,x)$ of water molecules' displacements $x(t)$ in each imaging voxel \citep{jones2010book}. With typical displacements $\sqrt{\langle x^2(t)\rangle}\sim 10\,\mu$m during  diffusion times $t\sim 50\,$ms used in the clinic, the dMRI signal becomes uniquely sensitive to how tissue structure on the micrometer scale restricts the diffusion of water molecules, opening a window into cellular-level details such as  cell density, shape, orientation, and permeability of cell membranes \citep{novikov2019,alexander2019}. Thanks to this unique \textit{in vivo} contrast, dMRI is particularly promising in detecting microstructural changes related to developmental and pathological processes in the brain white matter (WM), including myelination and demyelination, axonal growth and axonal loss, pruning, beading, and inflammation \citep{horsfield2002}. 

The greatest technical challenge of clinical dMRI is to uncover the exact relationship between cellular-level features and the dMRI signal --- i.e., to make dMRI not just sensitive, but specific to tissue microstructure. This would turn an empirical diagnostic technique into a quantitative and reproducible scientific measurement paradigm enabling improved understanding of the changes that underlie development, aging and disease, and tracking of its progression. So far, widely adopted dMRI techniques, such as diffusion tensor imaging (DTI) \citep{basser1994} and diffusion kurtosis imaging (DKI) \citep{jensen2005}, offer ways to represent the dMRI signal $S(t,q)$ (the Fourier transform of the displacements probability density $S(t,x)$) as expansions up to $q^2$ and $q^4$, correspondingly. However, these empirical signal representations inherently lack specificity to cellular-level phenomena, as they do not rely on any assumptions about tissue microgeometry.  

\begin{figure*}[b!!]
\centering
\includegraphics[width=0.85\textwidth]{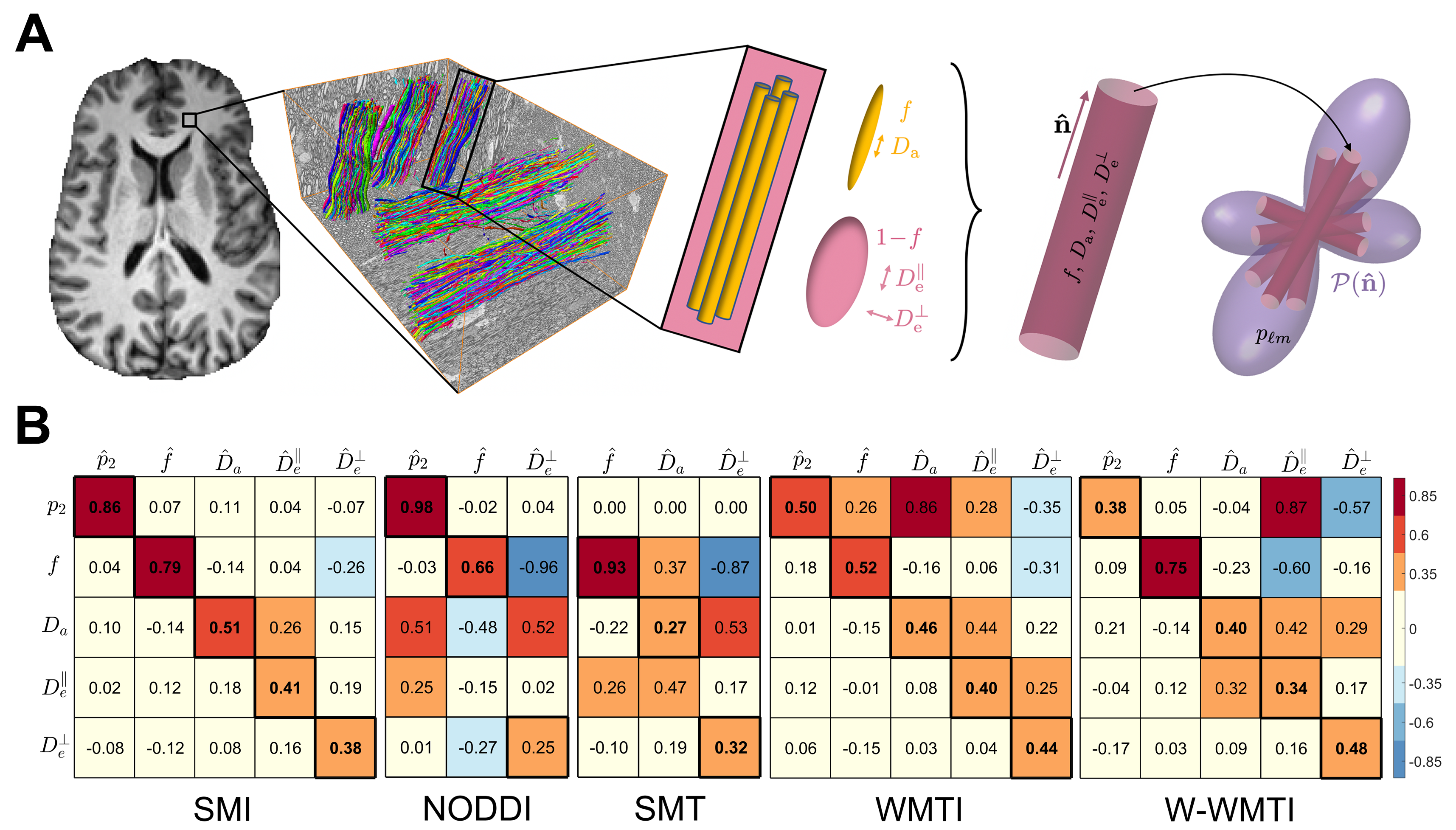}
\caption[caption FIG]{\textbf{Sensitivity and specificity matrix of SM estimators.} (\textbf{A}) Elementary fiber fascicles of the SM, consisting of the IAS and EAS, are described by at least 4 independent parameters: $f$, $D_a$, $D_e^{\parallel}$, $D_e^{\bot}$. The IAS is modeled as sticks with zero radial diffusivity, as axons are much narrower than the diffusion length of the dMRI measurement. Within a macroscopic imaging voxel, elementary fascicles contribute to the directional dMRI signal according to their orientation distribution function (ODF) $\mathcal{P}({\hat n})$. (\textbf{B}) The Sensitivity-Specificity Matrix (SSM) $S_{ij}$ is defined in Eq.~\ref{eq_SSM}. An ideal SSM is an identity matrix. The diagonal elements measuring sensitivity are in bold (some diagonal elements may not be on the diagonal line because the parameters that are not estimated are left out). The nonzero off-diagonal elements reveal spurious correlations between model parameters, and are a hallmark of decreased specificity. The SSM is color-coded to highlight the elements that are greater in absolute values.}
\label{fig:SSM}
\end{figure*}

In the pursuit of specificity, there has been a growing interest \citep{novikov2018modeling} in biophysical models that directly parameterize relevant tissue microgeometry and thus offer ways to quantify its changes in health and disease \citep{novikov2019,alexander2019,jelescu2017}. 
For WM, the {\it Standard Model} (SM) has been proposed as an overarching framework \citep{novikov2018,novikov2019,reisert2017}, unifying multicompartment model-based strategies over the past two decades  \citep{kroenke2004,jespersen2007,assaf2004,alexander2010,fieremans2011,zhang2012,kaden2016,reisert2017,novikov2018}, see Fig.~\ref{fig:SSM}A. In the SM, an elementary fiber fascicle is comprised of two non-exchanging compartments, the intra- and extra-axonal spaces (IAS and EAS). The SM offers an exciting potential of specificity to cellular-level biological phenomena, as its scalar parameters $f$, $D_a$, $D_e^{\parallel}$, $D_e^{\bot}$, as defined in Fig.~\ref{fig:SSM}A and described in detail below, are by design more specific to micrometer-level manifestations of pathological processes. 

\leftmarginnote{Address Comment R1.1} 
\commentNew{In the intracellular space, the axonal water fraction ($f$) characterizes the relative contributions of IAS and EAS water. A decrease in $f$ typically indicates axonal loss, suggesting a lower density of axons within the sampled voxel—a potential hallmark of neurodegeneration \citep{jelescu2016b}. The intra-axonal diffusivity ($D_a$) assesses axonal integrity. For instance, axonal injury, such as beading, disrupts the uniform diffusion pathway by introducing variations in axonal caliber, leading to the possible slowing or transient restriction of water molecules within axons \citep{budde2010}. Furthermore, within the EAS, the radial diffusivity ($D_e^{\bot}$) reflects the condition of the myelin sheath. The process of demyelination will reduce the complexity of pathways available to water molecules moving perpendicularly to the axons, resulting in increased $D_e^{\bot}$ \citep{jelescu2016b}. The axial diffusivity within the EAS ($D_e^{\parallel}$) detects a wider array of extra-axonal alterations. Decreases in $D_e^{\parallel}$ could indicate pathological events such as astrogliosis or microglial activation, both associated with neuroinflammatory responses \citep{xie2010rostrocaudal}} 

The practical relevance of the SM is rooted in its assumptions that make it compatible with clinically feasible diffusion acquisitions as well as presence of a large number of publicly available dMRI datasets with multiple diffusion weightings $b=q^2 t$ up to 2-3 \unit{ms/\mu m^2}, such as UK Biobank \citep{miller2016}, Human Connectome Project \citep{glasser2016}, Alzheimer's Disease Neuroimaging Initiative \citep{jack2008} and Adolescent Brain Cognitive Development \citep{casey2018}, together comprising hundreds of thousands of subjects. The application of the SM to these datasets presents unparalleled opportunities for large-scale \textit{in vivo} studies of tissue microstructure in health and disease. 

However, SM parameter estimation has proven to be challenging due to ``shallow" (almost degenerate) directions in the likelihood landscape \citep{jelescu2016,novikov2018}. \rightmarginnote{Address Comment R2.2} \commentNew{Conventional maximum likelihood estimation (MLE) suffers from low precision in this degenerate likelihood landscape. Recently, machine-learning (ML) based approaches have emerged as a promising tool for increased precision and speed. In the field of quantitative magnetic resonance imaging, ML have been used to estimate $\mathrm{T_1}$ and $\mathrm{T_2}$ \citep{cohen2018mr}, myelin water fraction \citep{liu2020myelin}, susceptibility \citep{yoon2018quantitative}, and dMRI model parameters \citep{golkov2016q,reisert2017,palombo2020sandi}. }

Here we quantify the sensitivity and specificity of ML-based SM estimation for relatively short ($\sim$6 min), clinically feasible dMRI protocols. Using dMRI acquired on patients during routine brain scans, we demonstrate how SM parameters are able to capture specific cellular-level changes in early development, stroke and multiple sclerosis. Our results open the way to apply this modern methodology in clinical settings and to large imaging consortium data \citep{miller2016,glasser2016,jack2008,casey2018} for investigating development, aging and pathology. 

\section{Theory}
\subsection{Assumptions of the Standard Model}
According to the SM, the dMRI signal originates from a collection of identical fiber fascicles in a WM voxel, that are oriented based on an arbitrary orientation density function (ODF) ${\cal P}({\hat n})$, Fig.~\ref{fig:SSM}A. The following SM assumptions specify the physics of diffusion inside an elementary fascicle. 
\begin{itemize} 
\item First, the signal from a fascicle is a sum of contributions from non-exchanging spin populations in the IAS and EAS. Water exchange can be neglected since myelin layers form a practically impermeable boundary for the relevant diffusion times.  
\item Second, the fascicle's IAS is represented as a collection of aligned zero-radius cylinders (``sticks"), such that diffusion occurs only along the stick, while the transverse diffusion is negligible since axonal diameters of $\sim 1\,\mu$m are much smaller than typical diffusion displacements in a clinical MRI measurement. The bulk along-stick diffusion coefficient $D_a$ is reduced relative to that of free water $D_0=3\,\unit{\mu m^2/ms}$ due to intra-axonal organelles and micro-variations of axonal shape, such as beads \citep{lee2020combio}.
(Diffusivities and $b$-values are in the units of \unit{\mu m^2/ms} and \unit{ms/\mu m^2} throughout this work.)
\item Third, diffusion in the fascicle's EAS is assumed to be anisotropic and Gaussian, characterized by the axially-symmetric tensor with parallel and perpendicular eigenvalues $D_e^{\parallel}$ and $D_e^{\bot}$. This means that diffusion at clinical diffusion times is assumed to be in the long-time limit, and any residual diffusion time-dependence \citep{novikov2014} is neglected. 
\end{itemize}

\rightmarginnote{Address Comment R2.7}
\commentNew{This ``impermeable stick" assumption has been verified in vivo in human WM by observing the distinct functional form at strong diffusion weightings that is inherent to such stick compartment \citep{mckinnon2017dependence, veraart2019scaling}.} The two SM compartments, IAS and EAS, define a response kernel for a fiber fascicle, which is a local bundle of aligned sticks with the extra-neurite space surrounding them. The kernel's signal is:
\begin{equation}\label{eq_SM}
\mathcal{K}\left(b,\xi \right)=s_0{\ }\left[fe^{-bD_a{\xi }^{2}}+\left(1-f\right)e^{-bD^{{\parallel }}_e{\xi }^{{2}}-bD^{{\bot }}_e\left(1-{\xi }^{2}\right)}\right]
\end{equation}
where $\xi \mathrm{=}\hat{g}\mathrm{\cdot }\hat{n}$ is the scalar product between the symmetry axis $\hat{n}$ of the kernel and the gradient direction $\hat{g}$. Further compartments, such as isotropic cerebrospinal fluid (CSF), $\sim f_{iso}e^{-bD_0}$, can in principle be added. However, given typical multi-shell  protocols with moderate $b$, the CSF fraction $f_{iso}$, which is a lot smaller than the intra axonal water fraction $f$ in WM, is very difficult to estimate. Fig.~\ref{fig:fw} shows that multi-shell protocols are not sensitive enough to estimate $f_{iso}$ at realistic SNR. Introducing the free-water compartment will further increase the difficulty of estimating other SM parameters, especially EAS diffusivities (as the EAS signal is similar in its functional form to that of the CSF). Therefore, in this work, we use a two-compartment kernel (\ref{eq_SM}) without CSF.

Such multicompartment fascicles are distributed in a voxel based on the fiber ODF. All fascicles are assumed to have the same compartment fractions and diffusivities, and differ from each other only by their orientation \citep{christiaens2020need}. Thus, the SM signal, measured along gradient direction $\hat{g}$, is a convolution between fiber response kernel $\mathcal{K}\left(b,\hat{g}\mathrm{\cdot }\hat{n}\right)$ and the ODF
$\mathcal{P}\left(\hat{n}\right)$ on a unit sphere:
\begin{equation}\label{eq_conv}
S_{\hat{g}}\left(b\right)=  \int_{\lvert\hat{n}\rvert=1}{{\rm d}\hat{n}}\, \mathcal{P}\left(\hat{n}\right)\mathcal{K}\left(b,\hat{g}\cdot\hat{n}\right) \end{equation}
where the ODF $\mathcal{P}\left(\hat{n}\right)$ is normalized to $\int_{\lvert\hat{n}\rvert=1}{\mathcal{P}\left(\hat{n}\right)}{\rm d}\hat{n}=1$.

\subsection{Rotational invariants in the spherical harmonic basis}

We factorize the kernel from the ODF parameters in Eq.~(\ref{eq_conv}) using the spherical harmonic (SH) basis \citep{reisert2017,novikov2018,tournier2007robust}:
\begin{equation}\label{eq_SH}
S_{lm}\left(b,x\right)=p_{lm}K_l(b,x) 
\end{equation}
where $S_{lm}$ and $p_{lm}$ are the SH coefficients of the signal $S_{\hat{g}}\left(b\right)$ and of the ODF
\begin{equation}\label{eq_ODF}
\mathcal{P}\left(\hat{n}\right)\approx 1+\sum^{l_{max}}_{l=2,4,\dots }{\sum^l_{m=-l}{p_{lm}}}Y_{lm}(\hat{n})
\end{equation}
up to order $l_{max}$ which practically depends on the dMRI sampling and signal-to-noise ratio (SNR). The functions $K_l\left(b\right)$ are projections of the kernel response onto the Legendre polynomials $P_{l}(\xi )$:
\begin{equation}\label{eq_legendre}
K_l(b)\equiv \int^1_0{{\rm d}\xi\, \mathcal{K}\left(b,\xi \right){P}_{l}(\xi )} \,.
\end{equation}

To factor out the dependence on the choice of the physical basis in three-dimensional space (via $m=-l\dots l)$, the rotational invariants are defined as follows \citep{reisert2017,novikov2018}:
\begin{equation}\label{eq_rotinv}
\begin{aligned}
S^{2}_{l}(b)&=\frac{1}{4\pi (2l+1)}\sum^{l}_{m=-l}\lvert S_{lm}(b)\rvert^{2} \\
p^{2}_{l}&=\frac{1}{4\pi (2l+1)}\sum^{l}_{m=-l}\lvert p_{lm}\rvert^{2}
\end{aligned}
\end{equation}

From the relationship between rotational invariants and SH coefficients, one can relate the rotational invariants to the kernel parameters \citep{reisert2017,novikov2018}:
\begin{equation}\label{eq_Sl}
S_l\left(b\right)=p_l K_l\left(b\right),\quad 
l=0,\ 2,\ 4, \ \dots\, .
\end{equation}
This enables a compression of raw directional dMRI measurements $S_{\hat{g}}\left(b\right)$ to a small number of data features $S_l$ without loss of information. Here, $p_0\equiv 1$ under the ODF normalization; the remaining ODF invariants, one for each $l$, characterize its anisotropy, with the normalization factor chosen so that $0<p_l<1$. Among these $p_l,\;l=2,4,6...$, $p_2$ has the lowest order and thus highest SNR. Combining $p_l$ of the ODF with the kernel parameters, the SM parameters of interest are defined as $\theta =\left\{f,D_a,D^{\parallel }_e,D^{\bot }_e; p_2, p_4, \dots\right\}$. We will focus on $p_2$, as the most easily interpretable ODF alignment metric.  

\subsection{Degeneracy of the estimation landscape}

For any diffusion direction, the SM signal is a sum of decaying exponentials. Parameter estimation for models of such kind is generally ill-posed. Specific near-degenerate dimensions in the MLE landscape, have been established for SM numerically \citep{jelescu2016} and analytically \citep{novikov2018}. In such a shallow MLE landscape, different combinations of model parameters may become indistinguishable in the presence of realistic noise, causing unstable estimation results. 

To improve the fit robustness with limited dMRI data, conventional maximum likelihood estimators apply constraints. For instance, Neurite Orientation Dispersion and Density Imaging (NODDI) \citep{zhang2012} and Spherical Mean Technique (SMT) \citep{kaden2016} both assume $D_{a}=D_{e}^{\parallel}$, with NODDI further fixing $D_{a}=D_{e}^{\parallel}=1.7 \unit{\mu m^2/ms}$. Both NODDI and SMT use a tortuosity model to derive $D_e^{\bot}$: $D_e^{\bot} = D_{e}^{\parallel} \cdot (1-f)$ \citep{szafer1995}. On the other hand, White Matter Tract Integrity (WMTI) \citep{fieremans2011} and Watson-WMTI (W-WMTI) \citep{jespersen2018} imply specific fiber ODF shapes and impose a square-root branch choice, $D_a\leqslant D_e^{\parallel}$ for WMTI and $D_{a}\geqslant D_e^{\parallel}$ for W-WMTI. These overly simplified constraints may introduce biases and thus result in spurious findings \citep{jelescu2015,jelescu2016,novikov2018,lampinen2017}. Since NODDI fixes $D_a$ and $D_e^{\parallel}$ at 1.7 \unit{\mu m^2/ms}, and SMT uses spherical average signals of b-shells to factor out ODF ($p_2$), their results may be left blank in the following figures when comparing these estimators.

\subsection{Feature count of the rotational invariants}

Fundamentally, the number of independent parameters one can in principle determine is tied to the information content of the dMRI signal (the number of independent features accessible from data at a given noise level). The estimation of the fascicle and ODF parameters factorizes in the spherical harmonics basis \citep{reisert2017,novikov2018}, Eq.~(\ref{eq_Sl}), such that the number of independent scalar signal features $N_b N_l$ is a product of the number $N_b$ of the $b$-shells in the $q$-space, and the number $N_l$ of the independent {\it rotational invariants} $S_l(b)$ of the signal (constructed from its spherical harmonics $S_{lm}(b)$, Eq.~(\ref{eq_rotinv})) accessible at a given noise level. 

In this work, as well as in publicly available dMRI datasets \citep{miller2016,glasser2016,jack2008,casey2018}, all our acquisitions have $N_b=2$ shells, and we use $S_l(b)$ with $l=0,2,4$, such that $N_l=3$, yielding overall $2\times 3 =6$ independent measurements for the fascicle response. This exactly matches  the number of independent SM parameters $\theta = \{f, D_a, D_e^\parallel, D_e^\bot; p_2, p_4\}$ contributing at these $l$, Eq.~(\ref{eq_Sl}). Note that employing the invariant $S_4(b)$ is crucial, since the system (\ref{eq_Sl}) with $S_0(b)$ and $S_2(b)$ for two $b$-shells yields only $N_b N_l = 4$ independent measurements for 5 nonlinearly-coupled parameters $\theta = \{f, D_a, D_e^\parallel, D_e^\bot; p_2 \}$. Practically, the signal invariants $S_l(b)$ decrease quickly with $l$; fortunately, maps of $S_l(b)$ up to $l=4$, as shown in Fig.~\ref{fig:rotinv_l4}A, display clear WM structure. Fig.~\ref{fig:rotinv_l4}B-C further demonstrates that $S_4(b)$ remains informative for the acquisitions discussed in this work.

\subsection{Unconstrained parameter estimation with machine learning}

As a faster alternative to MLE, a machine learning (ML)-based estimator was proposed to directly map rotational invariants $S_l(b)$ to the SM parameters $\theta_i$ \citep{reisert2017}. The ML-based estimator uses a ``soft" prior, as the prior distribution (training sets) implicitly regularizes the estimation in the training process, instead of imposing hard constraints on the model. Here we use an extended version of this method, dubbed Standard Model Imaging (SMI) \citep{coelho2022}, applicable to multi-dimensional dMRI. SMI uses third-order polynomial regression to map $S_l(b)$ with $l=0,2,4$ to a set 
$\theta_i = \{f, D_a, D_e^\parallel, D_e^\bot; p_2, p_4\}$. The same Gaussian distribution of SM parameter set $\theta=\left\{f,D_a,D^{\parallel }_e,D^{\bot }_e;p_2\right\}$ is used for training the ML-based estimator throughout this study with mean [0.5, 2, 2, 0.7, 0.45] and variance [0.06, 1, 1, 0.1, 0.06].

The five aforementioned estimators (SMI, NODDI, SMT, WMTI and W-WMTI) effectively measure the same set of parameters under the SM framework, but adopt different constraints (as summarized in Table S\ref{table:WM_model}), therefore resulting in different outcomes and trends. It has become a crucial need to determine the most reliable (sensitive and specific) diffusion model estimator for routinely used multi-shell protocols, which will enable leveraging the enormous publicly available datasets. To evaluate the performance of SM estimators, below we propose a metric to quantify the sensitivity and specificity of parameter estimation, similar to the concept of the confusion matrix in classification. We then apply these estimators in various \textit{in vivo} datasets, including early development, acute ischemia and multiple sclerosis (total $N=821$), and compare them in light of the current knowledge of relevant (patho)physiological processes in the WM. 

\section{Methods and Materials} 
\subsection{Subjects and dMRI acquisition} 

We studied various datasets that included two-shell dMRI scans ranging from 5 to 7 minutes long, that were acquired on patients referred for routine clinical brain MRI in the department of Radiology at New York University (NYU) and Medical University of South Carolina, indicating the potential for clinical translation of the proposed methods. Institutional Internal Review Board approval with waiver of consent was obtained for retrospective study. 

\leftmarginnote{Address Comment R2.3}
\subsubsection{Early development}
For assessment of human development, brain MRIs were studied of 59 pediatric subjects (30 females) who underwent DKI imaging as part of a routine MRI exam under sedation at NYU School of Medicine from June 2009 to October 2010 \citep{paydar2014,jelescu2015}. The subjects ranged from 1 day to 2 years and 9 months in age, and all underwent brain MR imaging for non-neurological indications. All the included exams were interpreted as normal by fellowship-trained board-certified neuroradiologists, and were reevaluated by a board-certified pediatric neuroradiologist for normalcy prior to inclusion.

All pediatric subjects were scanned on a 1.5 T Siemens Avanto MRI scanner using a body coil for excitation and 8-channel head coil for reception \citep{paydar2014,jelescu2015}. Whole brain diffusion weighted data were acquired using twice refocused spin-echo, single shot echo planar imaging with 1 b = 0 image and along 30 diffusion encoding directions for b = 1, 2 \unit{ms/\mu m^2}. Other parameters included: TR/TE: 4500/96 \unit{ms}, matrix size: 82 $\mathrm{\times}$ 82; 28--34 slices (no gap); and voxel size of 2.2--2.7 $\mathrm{\times}$ 2.2--2.7 $\mathrm{\times}$ 4--5 \unit{mm^3}, 1 average, acquisition time approximately 5 minutes. 


\subsubsection{Ischemia}
For assessment of (sub)acute stroke \citep{hui2012}, clinical and MRI data was reviewed for consecutive patients at the Medical University of South Carolina who were admitted due to acute onset of neurological symptoms and were subsequently diagnosed with acute or subacute stroke in the middle cerebral artery territory as the cause for neurological impairments. A total of 28 patients admitted to this institution between August 2011 and February 2012 were included. These patients underwent MRI 7 hours to 3 weeks after symptom onset (82\% of the stroke patients were scanned within the first week of symptom onset). Patients with a history of brain neoplasm or intracranial hemorrhages were excluded from study.

The stroke patients were scanned on a 1.5 T Siemens Avanto MRI scanner \citep{hui2012}. Diffusion-weighted images were acquired with 3 b-values (1 b = 0 image; 1 and 2 \unit{ms/\mu m^2} along 30 diffusion encoding directions) using a vendor-supplied single-shot twice-refocused spin-echo echoplanar imaging sequence. Other imaging parameters were: slice thickness = 3 \unit{mm} (no gap), number of slices = 40, TR/TE = 5500/99 ms, field-of-view = 222 × 222 \unit{mm^2}, acquisition matrix = 74 × 74, image resolution = 3 × 3 \unit{mm^2}, acceleration factor 2, acquisition time approximately 7 minutes.

\subsubsection{Multiple sclerosis and healthy controls}
For assessment of multiple sclerosis (MS), we studied 177 subjects (age 48.47 $\mathrm{\pm}$ 9.78 years old, 119 females) identified with a clinical diagnosis of MS using the McDonald criteria \citep{polman2011} who were referred for MRI of the head at NYU Langone Health between November 2014 and June 2020. Within one year of the MRI, disability status was assessed using the Patient Determined Disease Steps (PDDS) questionnaire, a validated nine-point patient-reported metric of disease severity \citep{kister2013}. MS patients were separated into mild MS ($0\leqslant\mathrm{PDDS}\leqslant3$) and severe MS ($4\leqslant\mathrm{PDDS}\leqslant7$) based on the need of canes for walking. 

In total, 557 healthy controls (age 45.29 $\mathrm{\pm}$ 13.94 years old, 388 females) were selected from headache patients with normal brain MRI, and no history of neurological disorder. The subjects were referred for MRI of the head at NYU Langone Health. To compare with the MS patients, 177 of healthy subjects (age 48.47 $\mathrm{\pm}$ 9.76 years old, 119 females) matched by age and sex were chosen as controls. Moreover, 177 adults aged between 25 and 35 years old (age 30.28 $\mathrm{\pm}$ 2.91 years old, 94 females) were selected out of the cohort to establish the normative values of SM parameters to compare with the pediatric population.  

Both MS patients and controls underwent clinically indicated MRI on a 3 T Siemens Magnetom Prisma (46.3\%) or Skyra (53.7\%) scanner. The dMRI protocol included a monopolar EPI sequence as follows: 4--5 b = 0 images, b = 1 \unit{ms/\mu m^2} along 20 directions and b = 2 \unit{ms/\mu m^2} along 60 directions, with imaging parameters: 50 slices, 130 $\mathrm{\times}$ 130 matrix, voxel size = 1.7 $\mathrm{\times}$ 1.7 $\mathrm{\times}$ 3 \unit{mm}, TE = 70--96 \unit{ms} and TR = 3200--4000 \unit{ms} on Prisma, TE = 95--100 \unit{ms} and TR = 3500--4300 \unit{ms} on Skyra, GRAPPA acceleration 2, and multiband 2. The total acquisition time is approximately 6 minutes.

\subsection{dMRI processing}
The dMRI data was processed using DESIGNER \citep{ades2018} for denoising \citep{veraart2016}, Gibbs artifact correction \citep{lee2021}, EPI-induced distortion correction \citep{andersson2003}, motion and eddy current artifact correction \citep{smith2004} and Rician noise floor correction \citep{koay2009}. Regions of interest (ROI) were automatically segmented by a nonlinear mapping onto the WM label atlas of Johns Hopkins University (JHU) \citep{mori2005}. \leftmarginnote{Address Comment R2.4} \commentNew{To mitigate the partial volume effects, we shrink each WM region by 1 voxel relative to the 1mm template.} The genu of corpus callosum (GCC) was chosen as the region of interest (ROI) for the MS study because it is a large homogeneous region in the corpus callosum with relatively few outliers.  \leftmarginnote{Address Comment R2.7} \commentNew{MS lesions, identified using icometrix \citep{rakic2021icobrain}, are notably heterogeneous \citep{rovaris2005diffusion} with potential exchange between the IAS and EAS in the case of unmyelinated axons and possibly additional compartments due to increased inflammation. Characterizing MS lesions is beyond the scope of this study. Hence, we focus on comparing the normal appearing white matter (NAWM) in the MS subjects with that of healthy controls.} For the stroke patients, the WM mask was determined by fractional anisotropy greater than 0.2 to include more voxels to the ROI for small ischemic lesions. 

SM parameters were estimated using the five WM estimators described, i.e. SMI, NODDI, SMT, WMTI, W-WMTI. The mean of an ROI was extracted for further analysis after excluding outliers. The voxels with unphysical SM parameter values were first excluded, then parameter values $\pm2\sigma$ away from the ROI mean were considered as outliers, where $\sigma$ is the standard deviation within an ROI. Typically, fewer than 5\% of the voxels are discarded.

\begin{figure*}[htbp]
\centering
\includegraphics[width=\textwidth]{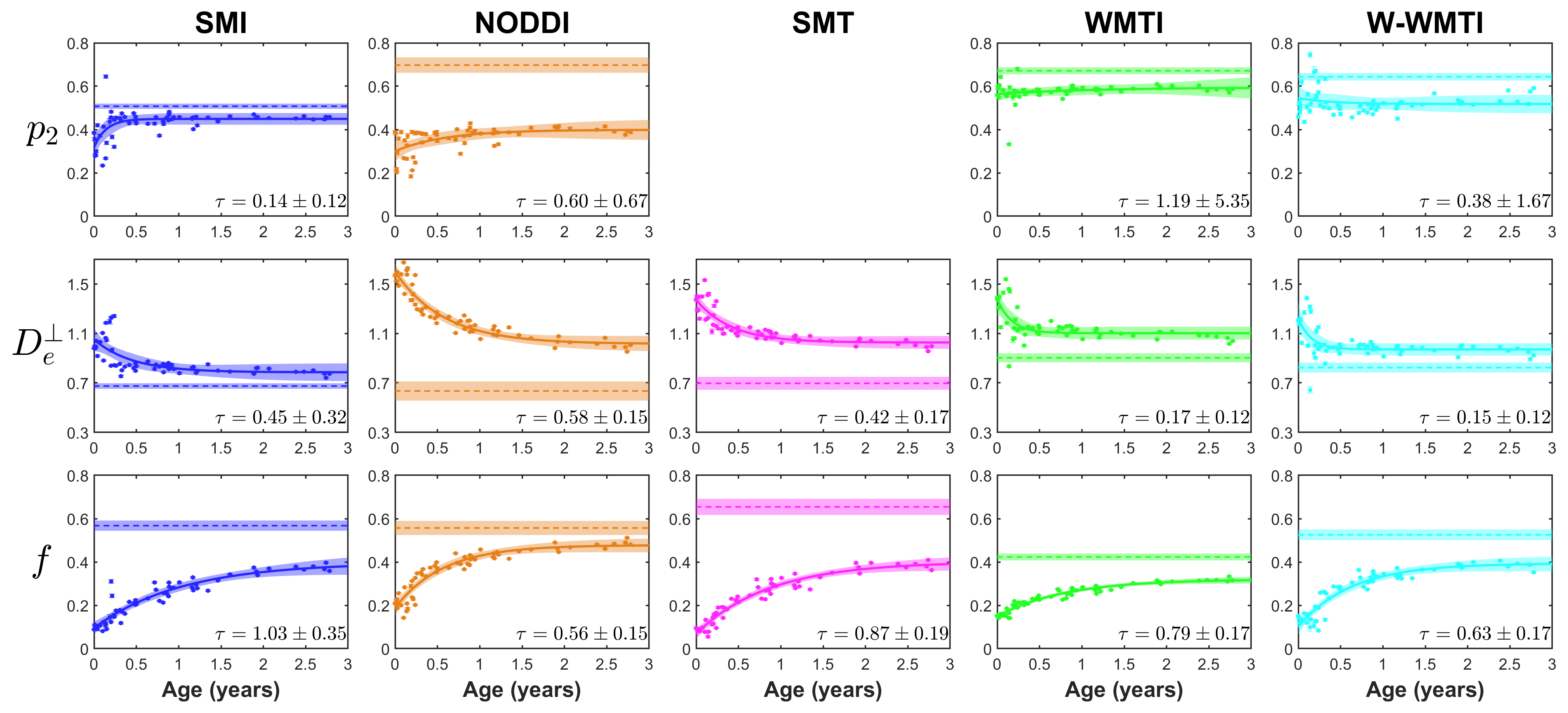}
\caption[caption FIG]{\textbf{Development trends of diffusion parameters in WM.} Parameters are ordered by the pace of development processes (based on histology studies): pruning ($p_2$) occurs faster than myelination ($D_e^{\bot}$), followed by axonal growth ($f$). Dots represent the WM mean of the pediatric subjects and error bars indicate their 95\% confidence interval. The exponential fit of the development data is plotted as a solid line within the 95\% confidence interval, while the 95\% confidence interval of its time constant $\tau$ is indicated on each plot. As a reference for the pediatric data, the dashed line and its neighboring shaded area represent the mean and standard deviation of the WM mean from 177 controls aged between 25 and 35. }
\label{fig:dev}
\end{figure*}

\subsection{Statistical analysis}
An exponential function ($A\cdot exp(-t/\tau)+B$) was fitted to the dataset of early development. The absolute value of $\tau$ was used to quantify the pace of the exponential growth or decay. For the stroke patients, the mean of ischemic lesions and their contralateral hemisphere in the WM were compared pairwise. The relative changes of ischemic lesions from their contralateral regions were calculated to quantify the degree of change for each SM parameter. In the MS study, MS patients were separated into two group: mild MS patients, severe MS patients based on the PDDS score. ANCOVA was used to study the group difference between every two groups covarying for age.

\subsection{Sensitivity-Specificity Matrix}

To quantify sensitivity and spurious correlations of parameter estimation, we consider the 
{\it Sensitivity-Specificity Matrix} (SSM) in noise propagations 
\begin{equation}\label{eq_SSM}
S_{ij} = \frac{\langle\theta_i\rangle}{\langle\theta_j\rangle}
\left \langle \frac{\partial \hat{\theta}_j}{\partial {\theta}_i}\right \rangle,\quad i,j=1,\hdots,N_{\theta} 
\end{equation}
whose elements quantify relative changes of an estimated parameter $\hat\theta_j$ with respect to the actual change of parameter $\theta_i$. Here, the angular brackets denote averaging over the distribution of ground truths (the test set) of all $N_{\theta}$ parameters. The normalization by the mean values $\langle\theta_i\rangle$ is introduced for convenience, to make the off-diagonal elements dimensionless (and are redundant for the diagonal elements). Practically, we evaluate the SSM from a linear regression of the estimates $\hat{\theta}_j$ with respect to ground truths ${\theta }_i$ of all $N_{\theta}$ parameters in a test set. 

Likewise, a linear regression of the estimated parameters was applied to the prior mean of SMI to demonstrate the dependency of ML-based estimation on the prior. We can define a matrix quantifying such impact as
\begin{equation}\label{eq_prior}
P_{ij} = \frac{\langle\hat{\theta}_{i}\rangle}{\langle\hat{\theta}_{j}\rangle} \left \langle \frac{\partial \hat{\theta}_j}{\partial \mu_{\theta_{i}}}\right \rangle,
\end{equation}
where $\mu_{\theta_{i}}$ are the mean values of the prior distribution for each model parameter. While fixing the variance of the prior distribution of SM parameters at [0.06, 1, 1, 0.1, 0.06], the prior mean was varied from 90\% to 110\% of the reference [0.5, 2, 2, 0.7, 0.45] at the step of 2.5\% for each parameter separately.

The synthetic data was generated based on the two-compartment SM using a two-shell protocol (same as \textit{in vivo} controls). The ODF was simulated by spherical harmonics up to $l=4$. The ratio of $p_4$ to $p_2$ was set between 0.75 and 0.85 according to histology results \citep{lee2019}. Gaussian noise was added to the signal at SNR = 25 with respect to $\mathrm{b}_{0}$ images. To evaluate the SSM, the ground truth of SM parameters $\theta=\left\{f,D_a,D^{\parallel }_e,D^{\bot }_e;p_2\right\}$ was uniformly sampled $1e^4$ times from [0.3, 1.5, 1.5, 0.4, 0.3] and [0.8, 2.5, 2.5, 1, 0.8] to focus on the most probable parameter range in WM voxels. Note that $D_e^{\parallel}\geqslant D_e^{\bot}$ was enforced in our simulation given their definition.

 \leftmarginnote{Address Comment R2.3}
 \section{Results and discussion}
\subsection{Sensitivity and specificity of SM parameters in simulations}

Results in Fig.~\ref{fig:SSM}B show that SMI provides the most trustworthy estimates of SM parameters. In particular, it estimates $p_2$, $f$ and $D_a$ almost free of spurious correlations. On the other hand, the SSM (Fig.~\ref{fig:SSM}B) reveals that NODDI parameters have non-negligible spurious correlations with other SM parameters, notably SSM$(D_{a},\hat{p}_{2})=0.51$ and SSM$(D_{a},\hat{f})=-0.48$. These spurious correlations are particularly concerning in the case of a significant change in $D_a$, e.g., in pathology, that would translate into apparent changes of $f$ and $p_2$. The estimation of $D_a$ by SMT has a combination of contrasts from $D_a$, $D_e^{\parallel}$ and $f$, and the estimation of $D_a$ by WMTI has a combination of contrasts of $D_a$ and $p_2$. These spurious correlations revealed by the SSM are caused by limited information obtainable from multi-shell dMRI scans and imposing hard constraints on the SM.

We show in Fig. \ref{fig:SP} the relationship between SM estimates and the prior mean given the same variance, which suggests the lower the sensitivity and specificity are, the more strongly the estimates are influenced by the prior distribution (training set). Their relationship is close to linear at the realistic SNR of dMRI (Fig. \ref{fig:SP}). Furthermore, the scatter plots of estimated parameters against ground truth for numerical noise propagations are shown in Fig. \ref{fig:scatterplots}A for all five estimators.

\rightmarginnote{Address Comment R3.1}
\commentNew{Specificity of the SM parameters has also been validated through histological studies. \cite{coronado2023volume} recently performed in a rat model of chronic traumatic brain injury an extensive comparison of diffusion MRI derived SM parameters against histology, evaluating different models including SMI and NODDI. They has demonstrated that the SM parameters for fiber dispersion are in excellent agreement with those derived from 3d electron microscopic images. Furthermore, the intra-axonal diffusivity agrees with the estimate from histology (based on the variation in the axon diameter). This work provides robust validation for SM parameters and demonstrate their specificity to geometric microscopic properties of WM tissues.}

\leftmarginnote{Address Comment R1.1}
\commentNew{SM parameters have been demonstrated to be sensitive to various WM processes. Notably, \cite{jelescu2016b} reported that demyelination leads to a decrease in $f$ and an increase in $D_e^{\bot}$, in a mouse model of de- and remyelination, suggesting there is no one-to-one correspondence between these two SM parameters, and both myelination and demyelination can influence $f$ and $D_e^{\bot}$ simultaneously. On the other hand, pruning will predominantly reduce the anisotropy of axon fibers ($p_2$) and to a lesser extent also might result in axonal loss ($f$). To maintain clarity in interpretation of the following SM findings, we focus on discussing the primary effects of specific processes.}

\subsection{Disentanglement of sequential processes in early development}
 Pediatric subjects (N=59) aged between 0 and 3 years old were selected for the study of early development \citep{paydar2014,jelescu2015}. As a reference for the pediatric data, 177 healthy controls aged between 25 and 35 were also selected. More detailed parameter distribution for healthy controls in the WM are presented in Fig. \ref{fig:scatterplots}B. Fig.~\ref{fig:dev} shows the changes in the SM parameters of $p_2$, $D_e^{\bot}$ and $f$ during early development in WM. \leftmarginnote{Address Comment R2.5} \commentNew{The results by NODDI for newborns are consistent with previous studies using NODDI \citep{dean2017mapping, kunz2014}. \cite{dipiero2023applications} offers a comprehensive review of dMRI studies focusing on early development.} The overall trends of early human brain development are largely consistent across the five estimators for $p_2$, $D_e^{\bot}$ and $f$ (Fig.~\ref{fig:dev}A). Yet, they differ in the pace of developmental processes, which can be quantified by the time constant $\tau$ of an exponential functional form  $A\,\exp(-t/\tau)+B$ for its dynamics. \leftmarginnote{Address Comment R2.4} \commentNew{On the other hand, Fig. \ref{fig:dev_supplemental} shows that the compartmental diffusivities are remarkably stable across the early development up to adulthood. Potentially higher fluid content in early development would result in an overestimation in $D_e^{\parallel}$ and an under-estimation of $D_a$ by SMI, which is not the case here.} 

\rightmarginnote{Address Comment R1.1}
\commentNew{Pruning is generally completed earlier than when the bulk of myelination occurs, as evidenced by animal histology (e.g., rats \citep{gorgels1990}, cats \citep{remahl1982,berbel1988}, rhesus monkeys \citep{lamantia1990}) and human imaging studies \citep{natu2019apparent, cafiero2019concurrence, fletcher2021oligodendrogenesis}. Pruning removes redundant axon collaterals and synaptic connections, increasing the anisotropy of ODF ($p_2$). SMI stands out as the only estimator that captures the rapid pruning after birth, as it shows $p_2$ has the smallest time constant among three SM parameters $p_2$, $f$ and $D_e^{\bot}$.} 

\commentNew{Both axon diameter growth and myelination span extensively, persisting into adulthood \citep{miller2012prolonged}. Results from SMI indicate that $D_e^{\bot}$ and $f$ are characterized by different time constants and carry distinct information, as opposed to NODDI linking these two parameters by the tortuosity relation. Moreover, the ``impermeable stick" assumption of the SM might not hold for axons that are not yet myelinated. Thus, the markedly low axonal water fraction at birth likely indicates the fraction of impermeable, myelinated axons rather than all axons. Therefore, $f$ is more indicative of myelination, with SMI exhibiting a time constant of 1.03±0.35 years. Among all SM estimators, this is closet to the 1$\sim$2 years of time constant measured from myelin water fraction \citep{deoni2012investigating}.} 

\rightmarginnote{Address Comment R2.5}
\commentNew{Furthermore, regional variances of diffusion in early development have been explored in various studies. \cite{paydar2014} and \cite{jelescu2015} discussed the order of development in DKI,  WMTI and NODDI, respectively. Their findings are generally consistent with the estimations of SMI. Regarding the intra-axonal fraction ($f$) estimated by SMI, we observed the following time constants: corticospinal tract ($0.51\pm0.38$), splenium of corpus callosum ($0.53\pm0.16$), posterior limb of internal capsule ($1.06\pm0.34$), and genu of corpus callosum ($1.21\pm0.37$). These early development trends in individual WM regions conform to the neuroscience principle that WM matures in a posterior-to-anterior and inferior-to-superior manner \citep{colby2011}.}

\begin{figure*}[htbp]
\centering
\includegraphics[width=0.78\textwidth]{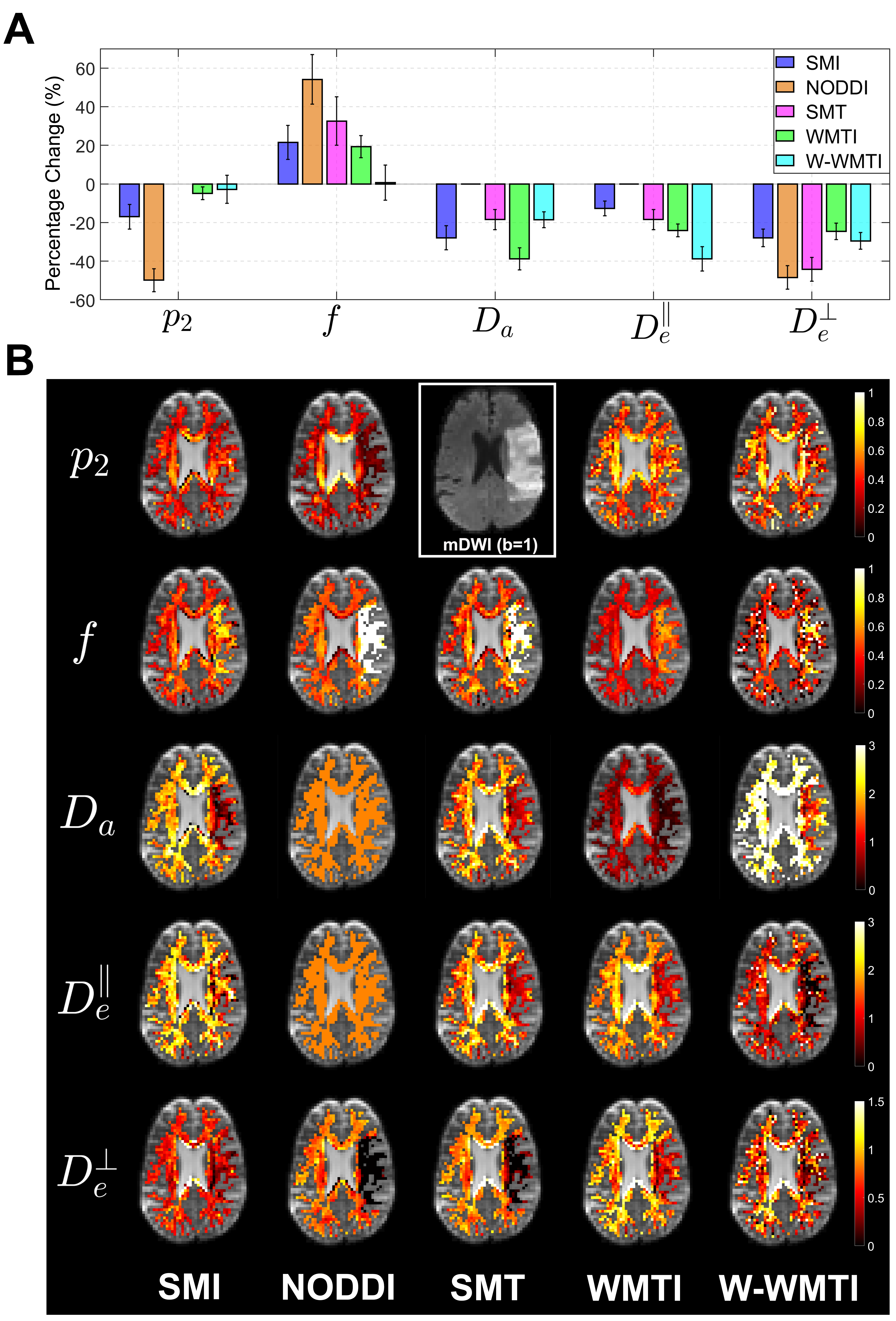}
\caption[caption FIG]{\textbf{White matter microstructure parameter changes in ischemic lesions as compared to the contralateral hemisphere.} (\textbf{A}) Mean relative percent changes of 28 subjects in diffusion parameters from normal (contralateral hemisphere) to (sub)acute ischemic tissues are shown in the bar graph (error bars indicate the 95\% confidence interval). (\textbf{B}) Parametric maps of WM are overlaid on the b=0 images of an exemplary stroke patient (scanned 26 hours after the onset of ischemia). In the middle of the top row (within the white box) is the mean DWI (mDWI) signal averaged over different directions of the $b=1$ \unit{\mu m^2/ms} shell, where the ischemic lesion is clearly shown.}
\label{fig:stroke}
\end{figure*}

\subsection{Detection of axonal beading in ischemic lesions}
Subjects (N=28) suffering from stroke and imaged with MRI from 6 hours to 2 weeks after ischemic onset were selected for the study of (sub)acute ischemia \citep{hui2012}. Fig.~\ref{fig:stroke}A shows the relative change of SM parameters in ischemic WM lesions compared to the same regions in the contralateral hemisphere. It is exemplified by the parametric maps of a stroke patient scanned 26 hours after the onset of ischemia in Fig.~\ref{fig:stroke}B. Remarkably, the $D_a$ map of SMI reveals that $D_a$ drops below 0.5 \unit{\mu m^2/ms} in a large portion of the ischemic region while it is around 1.5 \unit{\mu m^2/ms} in the contralateral hemisphere. In line with the parametric maps, the bar plots of SMI show $D_a$ and $D_e^{\bot}$ experience the largest decrease of $\sim 30\%$ averaged over all stroke patients, while $f$ increases by $\sim 20\%$ and $D_e^{\parallel}$ decreases by $\sim 15\%$. 

\leftmarginnote{Address Comment R1.2} \commentNew{Cytotoxic edema, commonly observed after a stroke, arises from the movement of water from the interstitial space into cells \citep{heo2005free}. This phenomenon occurs when extracellular Na$^+$ and other cations accumulate inside neurons and astrocytes, partly due to the breakdown of energy-dependent extrusion mechanisms \citep{liang2007cytotoxic}. dMRI is acutely sensitive to cerebral ischemia; the ADC from a dMRI scan drops dramatically within the infarcted region shortly after an ischemic event. From a microstructure perspective, \cite{budde2010} proposed that axonal beading could lead to a significant reduction in ADC, a hypothesis supported by simulations and in vitro experiments. Furthermore, \cite{novikov2014revealing} demonstrated that the structural disorder in the brain post-stroke is one-dimensional, based on diffusion time-dependence, a finding that corroborates the axonal beading model. In this context of axonal beading and cytotoxic edema, we anticipate the most profound change in intra-axonal diffusion being reduced due to beading ($D_a$), to a lesser extent an increase in axonal space ($f$) from the beading, and a reduction in the extra-cellular diffusion, axially ($D_e^{\parallel}$) and radially ($D_e^{\bot}$), due to extracellular constriction. SMI results are consistent with anticipated changes from axonal beading and cytotoxic edema, always falling within the SM parameter range. On the other hand, NODDI and SMT produce unphysical $f$ and $p_2$ values, indicating the assumptions $D_a=D_e^{\parallel}=1.7$ \unit{\mu m^2/ms} or $D_a=D_e^{\parallel}$ might not apply during ischemia.}

\begin{figure*}[htbp]
\centering
\includegraphics[width=0.7\textwidth]{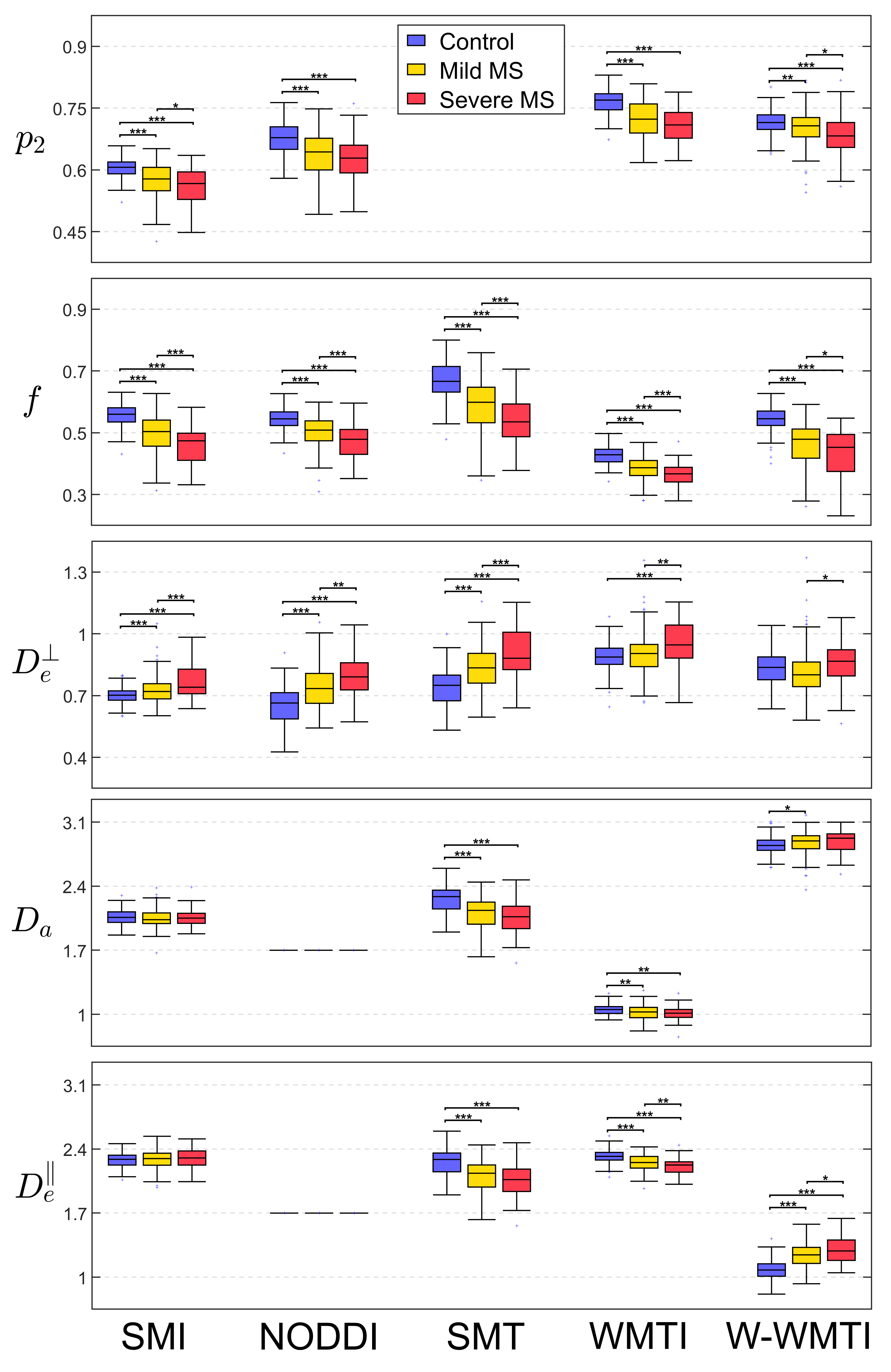}
\caption[caption FIG]{\textbf{Box plots of white matter microstructure parameters in MS patients and controls.} In comparison to controls (blue), MS patients are separated into mild MS (yellow) and severe MS (red). The mean of GCC is shown in the box plot after correcting for age. Please note that all diffusivities estimated by NODDI are constrained, i.e. $D_{a}=D_{e}^{\parallel}=1.7 \unit{\mu m^2/ms}$, $D_{e}^{\bot}=D_{e}^{\parallel} \cdot (1-f)$, while SMT assumes $D_a$ and $D_e^{\parallel}$ are identical but not fixed. On each box, the central mark indicates the median, and the bottom and top edges of the box indicate the 25th and 75th percentiles, respectively. The whiskers extend to the most extreme data points not considered outliers, and the outliers are plotted individually using the `+' marker symbol. The significance levels of the statistical ANCOVA test are displayed in asterisks on top of every two groups (*: p$<$.05; **: p$<$.01; ***: p$<$.001).} 
\label{fig:MS}
\end{figure*}

\subsection{Detection of axonal loss and demyelination in MS}
MS patients (N=177, 134 mild, 43 severe) were compared with 177 age and sex-matched controls in the GCC. \rightmarginnote{Address comment R1.4} MS lesions have been \commentNew{masked out} from the GCC before comparison. The severity of MS is determined based on Patient Determined Disease Steps (PDDS) questionnaire (mild: $0 \leqslant$ PDDS $\leqslant3$, severe: $4 \leqslant$ PDDS $\leqslant7$) \citep{kister2013}, where the clinical distinction between mild and severe MS is determined based on the ability to walk without a cane. The mean values of GCC normalized for age are shown in Fig.~\ref{fig:MS}. 

MS patients show lower $f$ and higher $D_e^{\bot}$ compared to controls, which is consistent with known MS pathology, i.e., axonal loss and demyelination \citep{trapp2008}. The decrease of $p_2$ in MS patients could be induced by the activation of microglia in the neuroinflammatory response \citep{voet2019}. An increasing number of microglia, which are morphologically plastic and considerably larger in size than axon diameters \citep{lawson1990}, may reduce the apparent anisotropy of ODF. While the five estimators detect largely consistent changes between controls and MS patients in $p_2$, $f$ and $D_e^{\bot}$, SMI is the only estimator that captures the continuous trend of MS severity from mild to severe cases in all three parameters. Nevertheless, the changes in $D_a$ and $D_e^{\parallel}$ are inconsistent among the five estimators (discussed below). 

\subsection{Source of discrepancies between SM estimators explained by the SSM}

Cellular and pathological specificity is the primary motivation for biophysical modeling of dMRI signals in brain WM. Yet, due to the limited information available in multi-shell dMRI protocols, biophysical models commonly employ model constraints to stabilize parameter estimation. These constraints tend to introduce unknown systematic biases and lead to discrepancies in parameter quantification and group comparisons. Despite efforts to compare the results of different WM estimators for the same dataset \citep{jelescu2015,beck2021}, the source of discrepancies so far has not been fully explained. We address this gap by proposing the SSM as a means to quantify the sensitivity and specificity of parameter estimation and thereby provide an explanation for the source of discrepancies in different estimators. The usefulness of SSM is illustrated here in various clinical datasets. 

In early development, the SSM suggests SMI has the highest specificity in estimating $p_2$. Indeed, SMI is the only SM estimator that captures the rapid axon elimination shortly after birth.

In ischemic lesions, the SSM predicts that SMI can capture the drop in $D_a$, while both SMT, and to an even larger extent NODDI, will instead cause the estimated $\hat{f}$ to increase, since NODDI completely fixes $D_a$ and SMT still allows $D_a$ to be estimated.

In MS, the primary pathological processes, axonal loss, demyelination and inflammation, affect $f$, $D_e^{\bot}$ and $p_2$ the most, whereas $D_a$ and $D_e^{\parallel}$ are less impacted. According to the results of SMI, there is no significant change of $D_a$ and $D_e^{\parallel}$ detected between controls and MS patients. Yet, the changes of $f$, $p_2$ and $D_e^{\bot}$ may ``leak" into the estimation of $D_a$ and $D_e^{\parallel}$. For instance, $\hat{D}_a$ estimated by SMT has a positive relationship with $f$ (SSM$(f,\hat{D}_a)=0.37$), which leads to the finding of a significant decrease in $D_a$. The same rationale can be used to further explain the other discrepancies in the MS results of $D_a$ and $D_e^{\parallel}$. 

\subsection{ML-based estimator vs MLE}
For the multi-shell dMRI protocols employed in clinical and large-scale studies \citep{miller2016,glasser2016,jack2008,casey2018}, the limited information necessitates employing constraints to stabilize MLE in a ``shallow" (almost degenerate) optimization landscape \citep{jelescu2016,novikov2018}. Constraints like $D_a=D_e^{\parallel}$ in SMT, or further $D_a=D_e^{\parallel}=1.7$ \unit{\mu m^2/ms} in NODDI significantly narrow down the shallow likelihood landscape and increase precision for $f$ and $p_2$, which is why these constrained estimators have been embraced. However, such constrained estimators introduce biases to the estimation and result in spurious findings \citep{jelescu2015,jelescu2016,novikov2018,lampinen2017}. WMTI takes a different approach by relating SM parameters to DKI metrics. To establish this analytically, WMTI assumes fiber alignment, essentially forcing $p_2$ close to 1. It is no surprise to see all of parameters estimated by WMTI are correlated with $p_2$ (Fig. \ref{fig:SSM}B). However, all SM parameters are not fixed throughout development, aging, and pathology. Hence, while an SM parameter can be fixed during the estimation process, the influence of a parameter may be reflected as changes in the other parameters, leading to spurious correlations. 

ML-based methods provide a promising alternative and hold some unique advantages over MLE, which usually relies on applying hard constraints to achieve robustness. First, they optimize the training error of all samples as a whole rather than one sample at a time as in MLE, which allows the ML-based estimator to learn the trade-off between bias and variance. Second, ML-based estimators can learn an individual mapping from signals to every parameter, whereas MLE can only search for the most probable combination of parameters. The learning approach enhances the sensitivity of estimators to parameters that are confounded by others. Third, ML-based estimators use a ``soft'' prior distribution as training sets to regularize the estimation instead of applying hard constraints. \rightmarginnote{Address Comment R2.8} \commentNew{The estimation of $p_2$ or $f$ by SMI does not depend on the diffusivity as evidenced by the lack of spurious correlations between them. This is because the ML model learns to estimate $p_2$ and $f$ through the training samples with varying diffusivities and is able to resolve them. In comparison, imposing hard constraints on diffusivities like in NODDI and SMT to stabilize the estimation process will inevitably lead to spurious correlations between the estimated $f$ and the ground truth of diffusivities, as shown in their SSM (Fig. \ref{fig:SSM}).} Fourth, SMI relies on rotational invariants up to $S_4(b)$, which is more informative than, say, SMT relying only on $S_0(b)$. 

\subsection{Limitations of the ML-based estimator}
By releasing hard constraints and using a ``soft'' prior in the training, SMI achieves higher sensitivity and specificity than conventional MLE algorithms. However, there are still several caveats. First, despite a large number of datasets available, only very limited information regarding the tissue microstructure of WM can be obtained from multi-shell protocols, which results in the spurious correlations revealed by the SSM. To further improve the parameter estimation of the SM, incorporating extra ``orthogonal" measurements is key in extracting complimentary information about the tissue microstructure for the ML algorithms to learn from. Planar and spherical diffusion encodings, and measuring dMRI signals at multiple echo times have been shown to significantly improve the estimation precision and accuracy \citep{topgaard2017,veraart2018,lampinen2020,coelho2022,reisert2019unique,coelho2019resolving}. 

\commentNew{Second, the SM may not be fully valid during early developmental stages or under certain pathological conditions.} \rightmarginnote{Address Comment R2.9} \commentNew{During very early development, there is minimal myelin around axons, potentially leading to water exchange between the IAS and EAS at the diffusion times employed in clinical diffusion MR protocols. We hypothesize that the SM parameter estimates might only measure myelinated axonal volume fractions when applied to developing WM. This hypothesis gains support from observations made from 0 to 3 years of age. During this period, there is a notable trend: the SM parameters—including $f$, $D_e^{\bot}$, $p_2$, gradually approach adult levels (Fig. \ref{fig:dev}). The consistent trajectories of SM parameters like $f$, $D_e^{\bot}$ and $p_2$ underscore the ongoing myelination process and other developmental activities. These findings suggest that SM parameters can effectively capture developmental changes.}

\leftmarginnote{Address Comment R2.1}
\commentNew{Third, since SMI is an ML-based estimator, it is inherently influenced by the choice of training sets. Figure \ref{fig:SP} illustrates that alterations to the prior mean induce a shift in the overall distribution. The amount of shift for a given parameter is the complement of its sensitivity (diagonal elements of the SSM). Thus, the impact of the prior distribution is more pronounced for parameters that are more difficult to estimate, such as compartmental diffusivities. As a result, the absolute values of estimates by SMI are only comparable using the same prior distribution. The ML-based estimator aims to minimize the mean squared error, which is a composite of both bias and variance. In this minimization process, a reduction in variance is attained at the expense of increased bias. This dynamic essentially nudges the estimates closer to the prior mean. The bias introduced in this manner is systematic, meaning it will not affect the outcome in group comparisons. To better control the prior mean, we use a Gaussian prior centered at a predefined value, with a cutoff at the physical bounds. The variance of the Gaussian prior mediates the bias-variance tradeoff inherent to the estimation process. A reduction in the variance of prior enhances the precision of estimates located at the center of the prior distribution, albeit potentially increasing the bias of estimates at the outer range of the prior. 

It’s crucial to understand that the bias observed for the ML-based estimation is not inherently a product of the prior itself. Instead, it emerges due to the insufficient information provided by the acquisition protocol and SNR, which are imperative for accurately resolving the parameters. As detailed in the supplement, the prior distributions adopted in this study do cover the entire physical range and are general enough to process various datasets. The mean of the prior distribution is substantiated by results from an extensive protocol with maximum b-value up to 10 \unit{ms/\mu m^2} \citep{novikov2018}. The variance of the prior distribution is selected to optimize the SSM for the multi-shell dMRI protocol at realistic SNR. This prior is recommended to be used in future studies with multi-shell dMRI protocols for comparison purposes.} 

\section{Conclusion}
In conclusion, because of the unique advantages enabled by ML-based estimators, as well as by relying upon the complementary rotational invariants $S_l(b)$ with $l=0,2,4$, SMI captures the biologically sensible morphological trends across three different clinical datasets, including early development, acute ischemia and MS. The SSM was proposed as a novel metric to measure the sensitivity and specificity of SM estimation, which largely explains the source of discrepancies between SM estimators and clearly demonstrates the highest sensitivity and specificity of SMI among five SM estimators of interest. Hence, SMI can serve as a powerful tool in clinical settings and for large imaging consortium data to study WM microstructure in a wide range of neurological diseases. 

\bibliographystyle{elsarticle-harv}
\bibliography{sn-bibliography.bib}

\section*{Acknowledgments}
The authors thank Sarah Milla for providing the dataset on early development, Edward S Hui, Jens Jensen and Joseph Helpern for providing the dataset of ischemia, Ilya Kister and Tamar Bacon for supplying clinical data and Wafaa Sweidan for organizing the dataset of multiple sclerosis.

\section*{Funding}
National Institute of Biomedical Imaging and Bioengineering grant P41EB017183 

National Institute of Neurological Disorders and Stroke grant R01NS088040 

National Institute of Biomedical Imaging and Bioengineering grant R01EB027075 

Irma T. Hirschl and Monique Weill-Caulier Trust 

National Institute of Health grants 1R01AG027852 and 1R01EB007656 

National Institute of Health and National Center for Research Resources grant UL1RR029882 

\section*{Author contributions}
E.F., D.S.N. designed and supervised research; Y.L. performed research and analyzed data; S.C. developed SMI code and contributed to data interpretation;  J.C., B.A. developed DESIGNER code and contributed to image processing; M.P., R.O., Y.W.L., T.S. characterized the clinical subjects; Y.L., E.F., D.S.N. wrote the manuscript.  All authors discussed the results and implications and commented on the manuscript at all stages.

\section*{Competing interests}
E.F., D.S.N. are co-inventors in technology related to this research with US patents US10360472B2 and US10698065B2.

\section*{Materials availability}
Code is available at https://github.com/NYU-DiffusionMRI/SMI.

\section*{Supplementary materials}
Figs. S1-S5

Table S1

\clearpage
\setcounter{equation}{0}
\pagenumbering{arabic} 

\newcommand{\beginsupplement}{
        \setcounter{table}{0}
        \renewcommand{\thetable}{S\arabic{table}}
        \setcounter{figure}{0}
        \renewcommand{\thefigure}{S\arabic{figure}}
        \setcounter{section}{0}
        \renewcommand{\thesection}{S\arabic{section}}
        \setcounter{equation}{0}
        \renewcommand{\theequation}{S\arabic{equation}}
     }
\beginsupplement

\begin{figure*}[htbp]
\centering
\includegraphics[width=\textwidth]{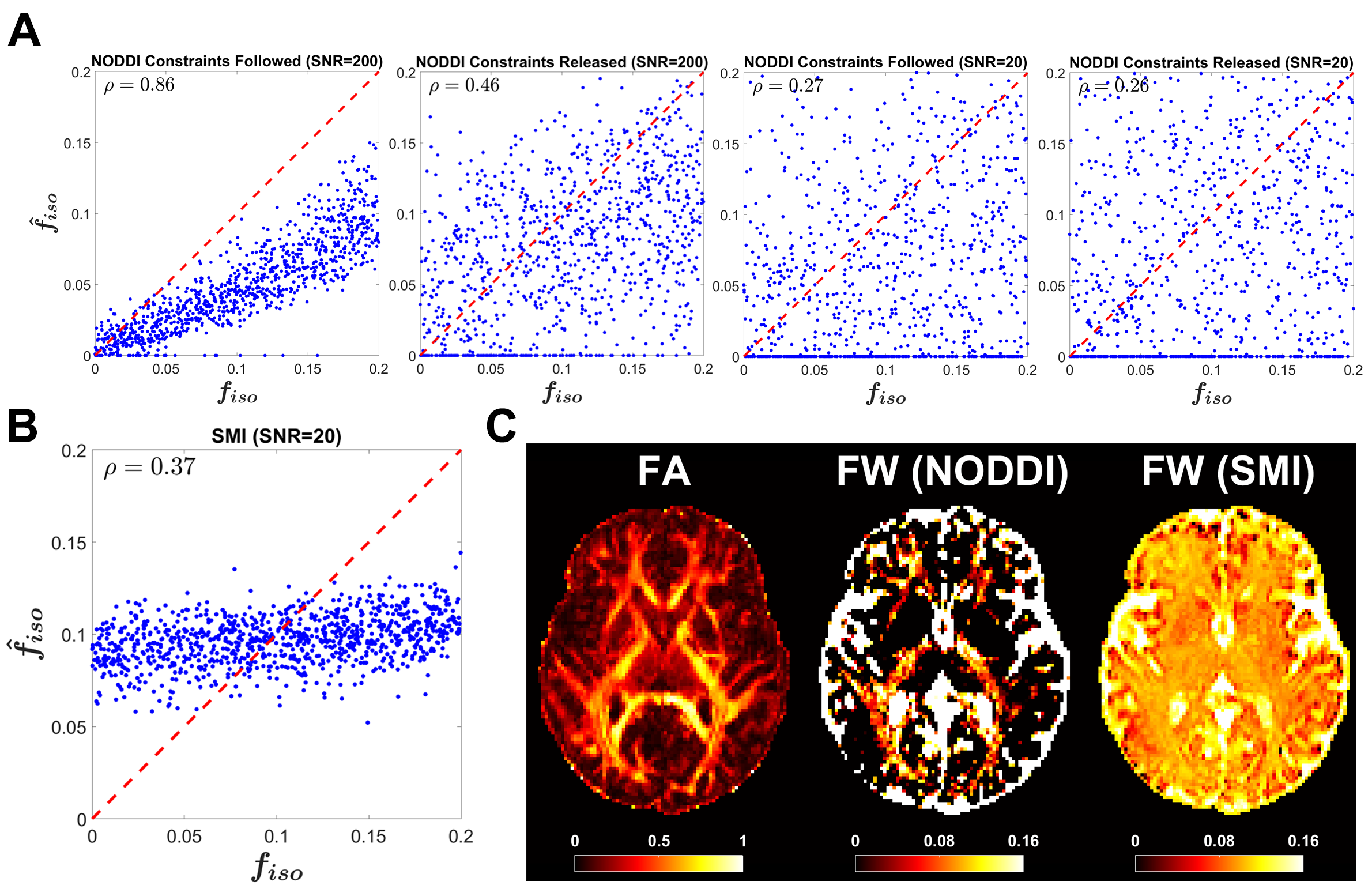}
\caption[caption FIG]{\textbf{Free water estimation with two-shell protocols using NODDI and SMI.} (\textbf{A}) NODDI estimate of free water volume fraction $\hat{f}_{iso}$ is plotted against the ground truth $f_{iso}$ of synthetic data for SNR=20 and 200, and for NODDI diffusivity constraints followed and released. Correlation $\rho$ between estimates and ground truth is indicated on each plot. (\textbf{B}) SMI estimate $\hat{f}_{iso}$ against the ground truth with SNR=20 and constraints released, same as the last plot in (\textbf{A}). The prior distribution of $f_{iso}$ is uniform between 0 and 0.2. As a result, the SMI estimate $\hat{f}_{iso}$ is roughly near the prior mean 0.1 due to lack of sensitivity to the CSF compartment.  (\textbf{C}) Exemplary parametric maps of a 43-year-old female control. Fractional anisotropy (FA) and free water volume fraction (FW) estimated by NODDI and SMI are presented. NODDI and SMI can distinguish fiber tracts and ventricles, but within a fiber tract, NODDI exhibits noisy estimates while SMI lacks contrast.}
\label{fig:fw}
\end{figure*}

\begin{figure*}[htbp]
\centering
\includegraphics[width=0.73\textwidth]{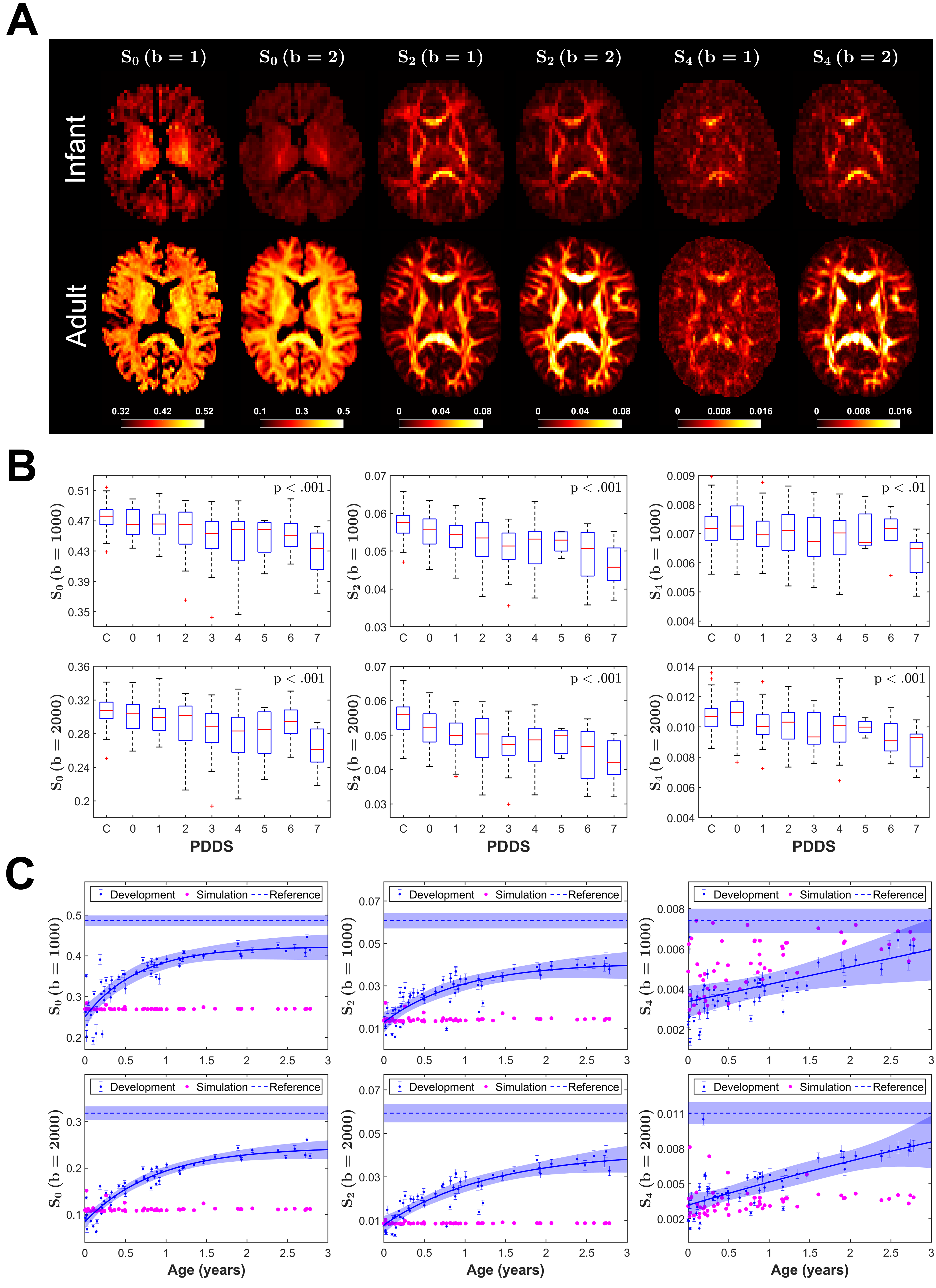}
\caption[caption FIG]{\textbf{Rotational invariants of the dMRI signals.} (\textbf{A}) Parametric maps of rotational invariants from a 3-month-old female infant and a 28-year-old female adult. b-value is in the unit of \unit{ms/\mu m^2}. (\textbf{B}) All rotational invariants of GCC mean excluding MS lesions (corrected for age) exhibit a significant decline (p-value indicated on the top right corner) as the severity of MS disability increases. On the leftmost position of the x-axis, `C' stands for controls. (\textbf{C}) To account for the SNR differences across subjects, the magenta dots represent the simulated rotational invariants of typical SM parameter combinations for newborn infants at the mean SNR level detected in each pediatric subject. In the trajectory of $S_4\,(b=1)$ alone, which has a lower SNR than the rest of rotational invariants as shown in (\textbf{A}), the effect of varying SNR overtakes the change caused by development.}
\label{fig:rotinv_l4}
\end{figure*}

\begin{figure*}[htbp]
\centering
\includegraphics[width=\textwidth]{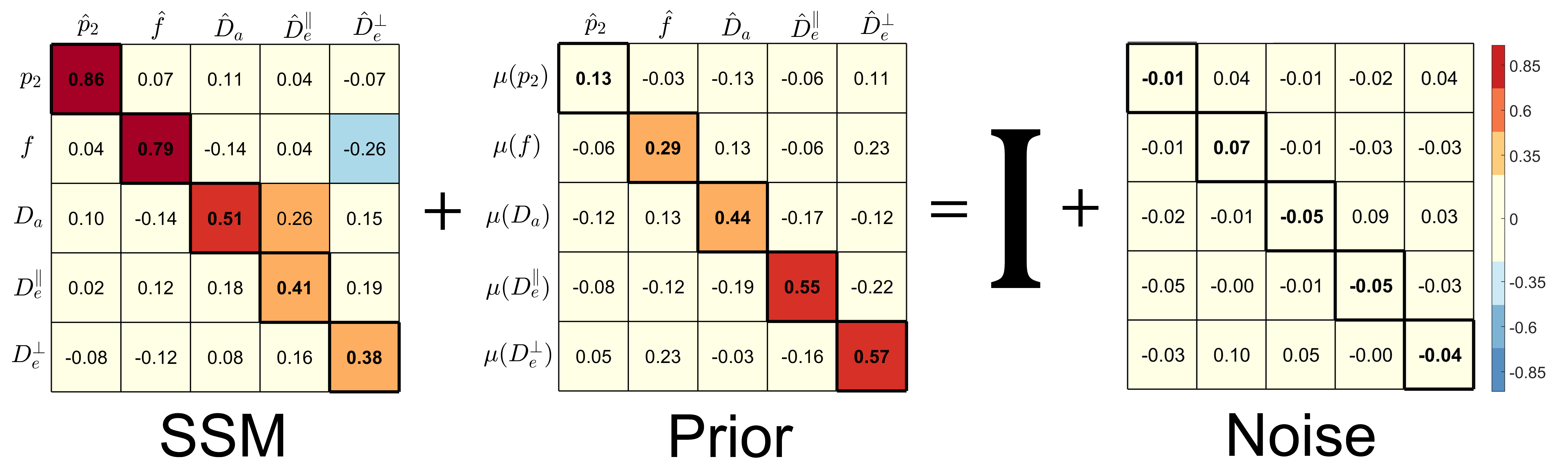}
\caption[caption FIG]{\textbf{Sensitivity-to-prior matrix.} Sensitivity-to-prior matrix was evaluated by applying linear regression to the prior mean of SMI to demonstrate the dependency of ML-based estimation on the prior. As is numerically validated in this figure, it is also possible to prove analytically that $S_{ij}+P_{ij} \approx I$ for a linear model estimated by a linear regressor, and this relationship becomes an approximation for a nonlinear model or regressor. This result suggests for ML-based estimators, deviations from the identity matrix in the SSM are related to the bias that is introduced to the estimator by the prior distribution (training set). }
\label{fig:SP}
\end{figure*}

\begin{figure*}[htbp]
\centering
\includegraphics[width=0.7\textwidth]{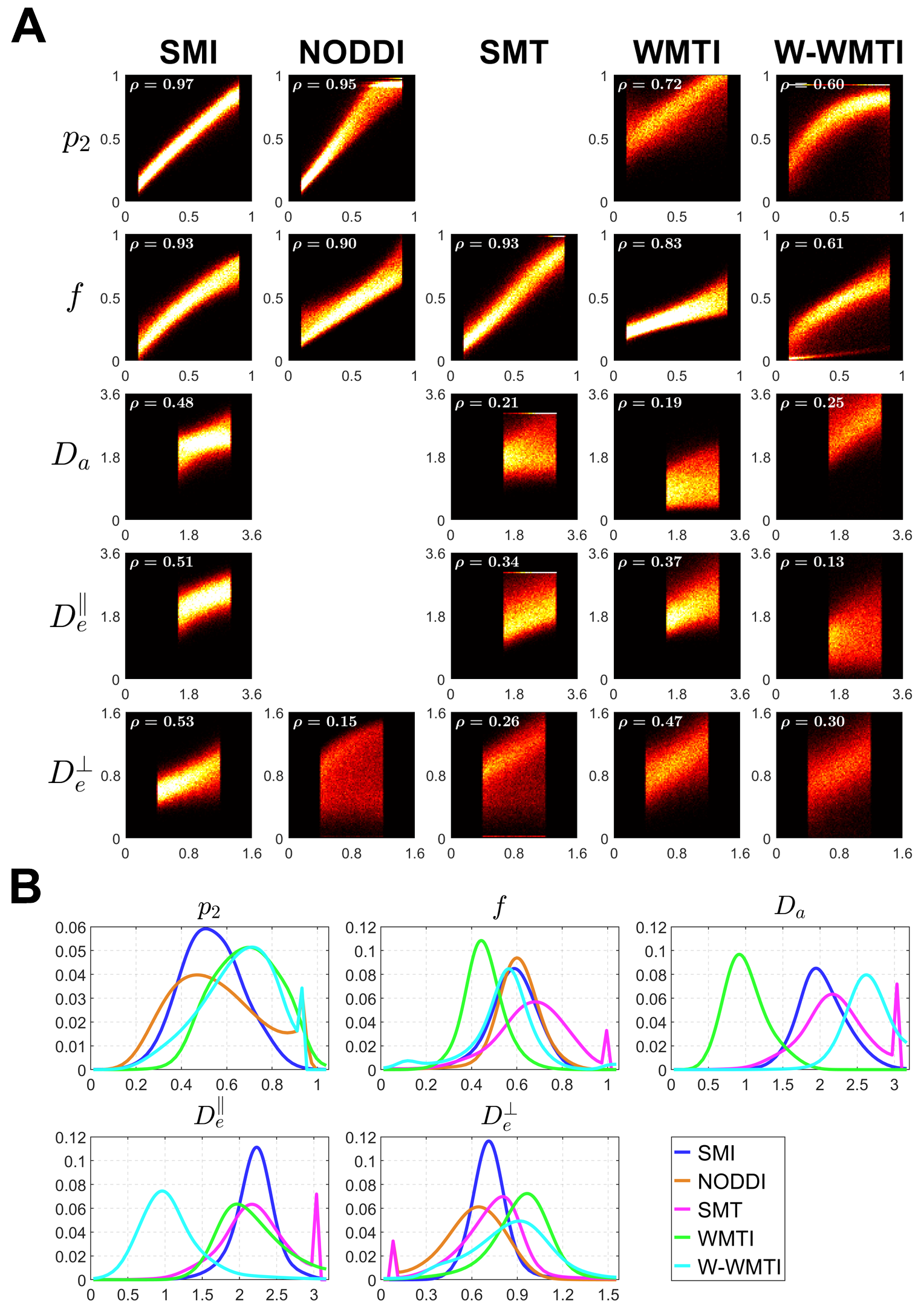}
\caption[caption FIG]{\textbf{Histograms of the SM parameter estimation.} (\textbf{A}) Simulations of realistic noise propagation for all estimators. Here, the estimates of each available SM parameter are plotted against the ground truth in scatter plots. Brighter color suggests higher data point density. Synthetic data are generated based on the SM for a two-shell protocol ($b = 1,2$) and realistic  signal-to-noise ratio SNR = 25 for $b=0$. The Pearson correlation coefficient $\rho$ between estimates and ground truth is indicated on each plot, where stronger correlations imply higher sensitivity. (\textbf{B}) Probability distributions of SM parameters from \textit{in vivo} data. Distributions are made up of over 200,000 WM voxels from 177 young controls aged between 25 and 35.}
\label{fig:scatterplots}
\end{figure*}

\begin{figure*}[htbp]
\centering
\includegraphics[width=\textwidth]{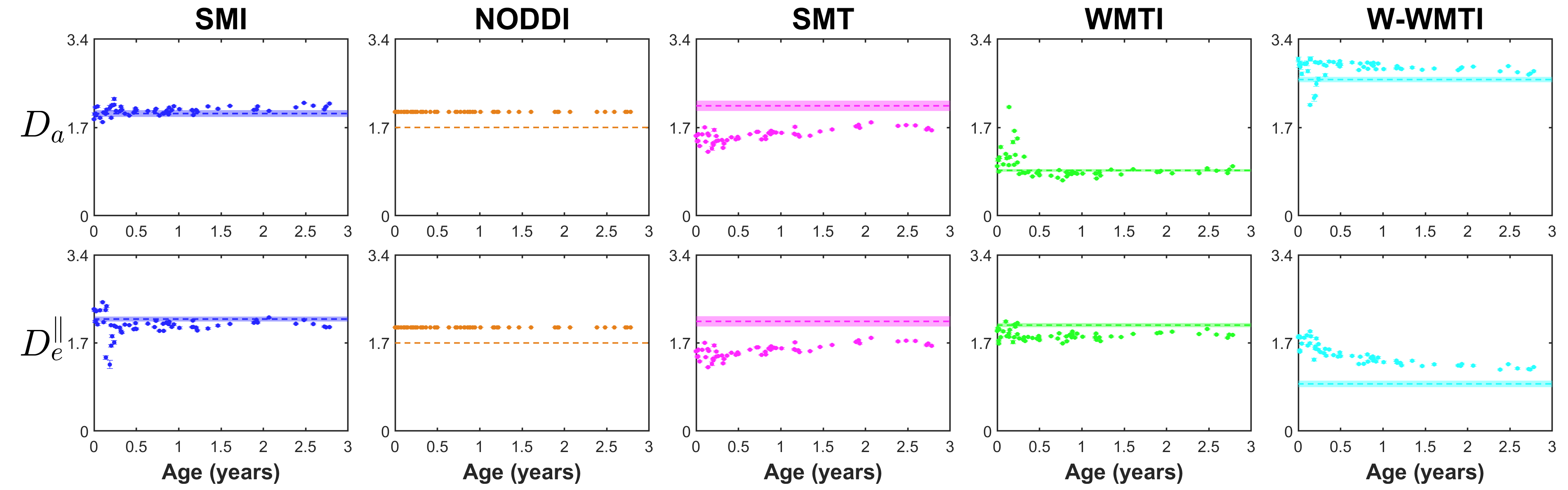}
\caption[caption FIG]{\textbf{Development trends of compartment axial diffusivities.} Early development trends between age 0 and 3. The dots represent the mean value of the entire WM for the pediatric subjects and the error bars indicate its 95\% confidence interval.  As a reference for the pediatric data, the dashed line and its neighboring shaded area represent the mean and standard deviation of the WM mean for 177 controls aged between 25 and 35. }
\label{fig:dev_supplemental}
\end{figure*}

\end{multicols}

\clearpage
\renewcommand{\arraystretch}{1}
\newcolumntype{C}[1]{>{\centering\arraybackslash}m{#1}}
\begin{landscape}
\begin{table*}[htbp]
\caption{Conceptual comparison between WM estimators} 
\centering 
\begin{tabular}{C{2.5cm}C{3cm}C{3cm}C{3.3cm}C{3cm}C{3.3cm}}
\toprule
 Estimator Name & Compartments & Diffusivity Constraints & ODF & Independent parameters & Estimation Method \\ [0.5ex]
\midrule
SMI & IAS + EAS & Unconstrained & Unconstrained & $p_2$, $f$, $D_a$, $D_e^{\parallel}$, $D_e^{\bot}$ & ML \\ 
NODDI & IAS + EAS + CSF & $D_{a} = D_e^{\parallel}=1.7$, $D_{e}^{\bot} = D_e^{\parallel}\cdot(1-f)$ & Watson distribution & $p_2$, $f$, $f_{iso}$ & MLE\\ 
SMT & IAS + EAS & $D_{a} = D_e^{\parallel}$, $D_{e}^{\bot} = D_e^{\parallel}\cdot(1-f)$  & Factored out by spherical mean & $f$, $D_a$ ($D_e^{\parallel}$) & MLE\\
WMTI & IAS + EAS & $D_{a}\leqslant D_e^{\parallel}$ & Fibers more or less aligned & $p_2$, $f$, $D_a$, $D_e^{\parallel}$, $D_e^{\bot}$ & Analytically derived from DKI metrics\\
W-WMTI & IAS + EAS & $D_{a}\geqslant D_e^{\parallel}$ & Watson distribution & $p_2$, $f$, $D_a$, $D_e^{\parallel}$, $D_e^{\bot}$ & Analytically derived from DKI metrics\\
\bottomrule 
\end{tabular}
\label{table:WM_model} 
\end{table*}
\end{landscape}

\end{document}